\documentclass[aps,prd,superscriptaddress,showpacs,showkeys]{revtex4}
\usepackage{amssymb}
\usepackage{amsmath}
\usepackage{amsfonts}
\usepackage{graphicx}
\usepackage{color}
\usepackage{changes}
\usepackage{enumerate}
\usepackage{changes}
\usepackage{palatino}
\usepackage[colorlinks=true,
            linkcolor=red,
            urlcolor=blue,
            citecolor=blue]{hyperref}
\newcommand{\beq}{\begin{equation}}
\newcommand{\eeq}{\end{equation}}
\newcommand{\bea}{\begin{eqnarray}}
\newcommand{\eea}{\end{eqnarray}}

\def\l{\left}
\def\r{\right}

\def\ql{\textquotedblleft}

\def\1o2{{1\over2}}
\def\e={\equiv}
\def\={\neq}
\def\>{\geq}
\def\<{\leq}

\def\b{\beta}

\def\g{\gamma}

\def\p{\phi}

\begin{document}
\title{ Effect of null aether field on weak deflection angle of black holes}

\author{A. \"{O}vg\"{u}n}
\email{ali.ovgun@emu.edu.tr}
\homepage{https://aovgun.weebly.com} 
\affiliation{Physics Department, Eastern Mediterranean University, Famagusta, North Cyprus, 99628 via Mersin 10, Turkey.}

\author{\.{I}. Sakall{\i}}
\email{izzet.sakalli@emu.edu.tr}
\affiliation{Physics Department, Eastern Mediterranean University, Famagusta, North Cyprus, 99628 via Mersin 10, Turkey.}

\author{J. Saavedra}
\email{joel.saavedra@ucv.cl}
\affiliation{Instituto de F\'{\i}sica, Pontificia Universidad Cat\'olica de Valpara\'{\i}%
so, Casilla 4950, Valpara\'{\i}so, Chile.}

\begin{abstract}
We study the light rays in a static and spherically symmetric gravitational field of null aether theory (NAT). To this end, we employ the Gauss-Bonnet theorem to compute the deflection angle by a NAT black hole in the weak limit approximation. Using the optical metrics of the NAT black hole, we first obtain the Gaussian curvature and then calculate the leading terms of the deflection angle. Our calculations show how gravitational lensing is affected by the NAT field. We also show once again that the bending of light stems from a global and topological effect.

\end{abstract}

\date{\today}
\maketitle

\section{Introduction}


According to the general relativity (GR) theory \cite{iz1}, gravity affects spacetime geometry, the gravitational field produced by matter can be powerful enough to drastically alter the ordinary causal structure of spacetime and produce a region, the so-called black hole (BH), in which even light is held and no part can escape to infinity. For this reason, BHs have been studying in detail by scientists over the years. 

The event horizon of a BH is a limit to which events cannot influence an observer on the opposite side. Namely, it is the gravitational point of no return that nobody can really "see" a BH. In fact, the event horizon is a null surface that separates the rays of light that reach infinity from those falling into singularity.On the other hand, astronomers can observe this curvature in spacetime as light from distant objects gets bent while going around foreground objects. In particular, if a BH is immersed in a bright region, like a disc of glowing gas, BH creates a dark region similar to a shadow \cite{izshadow,izshadow2}. This shadow, caused by the gravitational bending and capture of light by the event horizon, reveals a lot about the nature of these massive objects. From this information, the position and mass of the BH can be determined. For static and spherical symmetric spaces, suitable coordinate systems can be introduced to determine the location of the event horizon by looking at the places where the local light cones tilt over. This means that the existence of the event horizon (i.e., BH) is related to the local Lorentz invariance of time. Therefore, it is of great importance to study the properties of BHs in gravity theories that show local Lorentz invariance violations. Namely, in GR, the existence of BHs rides on the causal structure originated from both the Lorentz symmetry on matter fields and the local flatness theorem. Therefore, it is natural to examine whether BHs will still be formed in the absence of Lorentz symmetry or not; if they exits, it is even more important to analyze their physical features, since the signatures of Lorentz violations appear in the regime of strong gravity. On the other hand, when the Lorentz symmetry is broken and the causal structure is modified in a most radical way, BH solutions surprisingly do exist. In this new BH solutions, the event horizon is replaced by the universal horizon, which is to capture any mode independently of the propagation velocity \cite{iz2,iz3}. Some BH solutions have been already found in restricted Lorentz violating gauges, such as spherical symmetry \cite{iz4,iz2} and slowly-rotating background, both in lower dimensions \cite{iz6} and 4-dimensions \cite{iz7,iz8,iz9,iz10}. Generally, numerical solutions to the equations of Ho\v{r}ava gravity and Einstein-aether theory admit those BHs. Nevertheless, there are rarely analytical solutions for this type of theories. However, all those solutions are obtained in the symmetry-restricted scenarios for various asymptotics \cite{iz11}. In Einstein-aether theory, the vector field is timelike everywhere and explicitly breaks the boost sector of the Lorentz symmetry which has been studied extensively in the literature (see for instance \cite{iz12}). In particular, aether models present themselves as phenomenological probes to test for the presence of Lorentz symmetry breaking (LSB) in astrophysical objects and cosmology. Aether models are nothing but a vector input to the Lagrangian density of the system having a non-vanishing vacuum expectation. Because of this property, the vector field dynamically selects a opted frame at each point in spacetime and automatically breaks the Lorentz invariance. This is a mechanism that reminds the breaking of local symmetry in Higgs mechanism \cite{iz13} and serves as a phenomenological representation of the LSB terms in Standard Model Extension's gravitation sector \cite{iz14}.

NAT is one of the new vector-tensor theories of modified gravity theory \cite{iz15}. In this theory, the dynamical vector field acts as the aether and  exact spherically symmetric BH solution with charge is possible \cite{iz16}.  It was also discussed in \cite{iz16} that NAT fields have effects on the solar system dynamics when extracting the Eddington-Robertson-Schiff parameters $\b$ and $\g$, which appear in the perihelion precession and the light deflection expressions, for the NAT BHs. Furthermore, it was shown that at the post-Newtonian order although there is no contribution from the NAT field to the deflection of light rays passing near the BH (as in GR), however, NAT field gives contribution to the perihelion precession of planetary orbits. Later remark can play a role on solar system observations. The physical features (singularity structure, ADM mass, and thermodynamics) of the NAT BH are also analyzed in \cite{iz16}. Moreover, NAT charge is able to reduce the horizon thermodynamics to that of the Reissner-Nordstr\"{o}m-(A)dS BH of GR and modifies the circular orbits of massive and massless particles around the BH. 
To utilize the differential deflection exhibited by weak lensing, first weak deflection angle is calculated. It depends on the mass distribution of the gravitational lensing system. Gibbons and Werner showed that it is possible to calculate the deflection angle in weak field limits using the Gauss-Bonnet theorem (GBT) and the optical geometry \cite{R8,R5}. In this method, they focus the domain outside the trajectory of light. Optical metric has geodesics, which are the spatial light rays, that the focusing of light rays are considered as a topological effect \cite{Ovgun:2019wej}.  At the present time, the GW method has been applied to various spacetime metrics of black holes and wormholes (see for example \cite{Ovgun:2018xys,Jusufi:2017mav,Jusufi:2017vew,Sakalli:2017ewb,Jusufi:2017lsl,Ono:2017pie,Jusufi:2017hed,Jusufi:2017vta,Jusufi:2017uhh,Arakida:2017hrm,Crisnejo:2018uyn,plasma,Crisnejo:2019xtp,Jusufi:2018jof,Ovgun:2018fnk,Ovgun:2018ran,Ovgun:2018prw,Ono:2018ybw,Ovgun:2018oxk,Ovgun:2018tua,Ono:2018jrv,Jusufi:2018kmk,Ovgun:2018fte,Javed:2019qyg,Javed:2019a,Javed:2019b,Javed:2019jag,Javed:2019rrg,Kumaran:2019qqp,Li:2019vhp,GON1,GON2,GON3,GON4,GON5} and references therein). We can write the domain surface define as $(D, \chi, g)$ using the the Euler characteristic $\chi$ and a Riemannian metric $g$. Then the GBT can be formulated as follows \cite{R8}:
\begin{equation}
    \int \int_D K dS + \int_{\partial D} \kappa dt + \sum_i \alpha_i = 2\pi \chi(D) \label{GBT1}
\end{equation}
where, $\alpha_i$ is the exterior angle with $i$\textsuperscript{th} vertex, Gaussian curvature, $K$ and geodesics curvature, $\kappa$. This method, only works for the asymptotically flat spacetimes and it is given by \cite{R8}:
\begin{equation}
    \hat{\alpha} = - \int \int_D K dS. \label{GBT2}
\end{equation}

In this article, our main purpose is to explore the NAT effects on the gravitational lensing. To this end, we organize the paper as follows: In section II, we briefly review the BH spacetime of the NAT. Section III is devoted to computation of the deflection angle by NAT BH using the GBT in weak field regime and in the plasma medium. We conclude our results in section IV. Natural units are used throughout this paper: $\hbar=c=1$.

\section{NAT BH SPACETIME}

The line-element of the asymptotically flat NAT BH is given by \cite{iz16}
\begin{equation} \label{}
  ds^2=-h(r)dt^2+\frac{dr^2}{h(r)}+r^2d\theta^2+r^2\sin^2\theta d\varphi^2,
\end{equation}
where
\begin{equation}\label{}
  h(r)=\left\{\begin{array}{ll}
         \displaystyle 1-\frac{2a_1^2b_1}{r^{1+q}}-\frac{2a_2^2b_2}{r^{1-q}}-\frac{2\tilde{m}}{r}
         &\mbox{(for $q\neq0$),}\\
         &\\
         \displaystyle 1-\frac{2m}{r}
         &\mbox{(for $q=0$).}\label{iz0}
\end{array} \right.
\end{equation}

In the above equation, where $a_1$, $a_2$, $\tilde{m}$, and $m$ are just integration constants and
\begin{equation}\label{qbb}
  q\equiv\sqrt{9+8\frac{c_1}{c_{23}}},~~b_1=\frac{1}{8}[c_3-3c_2+c_{23}q],~~b_2=\frac{1}{8}[c_3-3c_2-c_{23}q],
\end{equation} 

in which $c_{1}$, $c_{2}$, $c_{3}$, and $c_{23}=c_{2}+c_{3}$ are the dimensionless constant parameters. Furthermore, the constants $\tilde{m}$ and $m$ are the mass parameters of the solutions.

As is obvious from Eq. (\ref{iz0}), in the $q=0$ case, the metric is nothing but the well-known Schwarzschild spacetime, which is asymptotically flat. However, in the $q\neq0$ case, to achieve asymptotically flat boundary conditions, one should consider the following cases separately (by definition $q>0$ \cite{iz16}):

\begin{equation}
 h(r)\mid_{r\rightarrow\infty}=1 \left\{\begin{array}{ll}
         \displaystyle  \mbox{for $0<q<1$}
         &\mbox{(if $a_1\neq0$ and $a_2\neq0$) or (if $a_1=0$ or $b_1=0$),}\\
         &\\
         \displaystyle  \mbox{for $0<q$}
         &\mbox{(if $a_2=0$ or $b_2=0$).}\label{hinf}
\end{array} \right.
\end{equation}

In this study, we shall consider the case of $a_2=0$. Then, the metric function $h(r)$ and the scalar aether field $\p(r)$ take the following forms
\begin{eqnarray}
&&h(r)=1-\frac{2a_1^2b_1} {r^{1+q}}-\frac{2\tilde{m}}{r},\label{hLa}\\
&&\p(r)=\frac{a_1}{r^{(1+q)/2}}.\label{phiLa}
\end{eqnarray}
The location of the event horizon $r_0$ is given by $h(r_0)=0$ and the area of the event horizon is $A=4\pi r_0^2$. Setting $a_1=GQr_0^{(q-1)/2}$, where $Q$ is the NAT \ql charge", Eqs. (\ref{hLa}) and (\ref{phiLa}) become
\begin{eqnarray}
&&h(r)=1-\frac{2G^2Q^2b_1} {r^2}\l(\frac{r_0}{r}\r)^{q-1}-\frac{2\tilde{m}}{r},\label{hLa1}\\
&&\p(r)=\frac{GQ}{r}\l(\frac{r_0}{r}\r)^{(q-1)/2}.\label{metric2}
\end{eqnarray}
At the location of $r_0$, we have
\begin{eqnarray}
&&h(r_0)=1-\frac{2G^2Q^2b_1} {r_0^2}-\frac{2\tilde{m}}{r_0}=0,\label{hLa2}\\
&&\p(r_0)=\frac{GQ}{r_0}.\label{phiLa2}
\end{eqnarray}
It is worth noting that the horizon condition (\ref{hLa2}) is independent of the parameter $q$. In addition, the scalar aether field $\p(r)$ resembles the electric potential at $r=r_0$.

Using the asymptotically flat solutions of NAT black hole given in \cite{iz16},taking $q=1$, the metric function and scalar aether field is

\begin{eqnarray}
&&h(r)=1-\frac{2a_1^2b_1} {r^2}-\frac{2\tilde{m}}{r},\label{metric}\\
&&\p(r)=\frac{a_1}{r^{1/2}}.
\end{eqnarray}

The deflection angle of photon can be calculated using the following formula ( $r_0$ is the distance of closest approach)  \cite{Virbhadra:1998dy}:
\begin{equation}
  \hat{\alpha}(r_0) = -\pi + 2 \int_{r_0}^\infty dr \frac{1}{r \sqrt{\frac{r^2 f(r_0)}{r_0^2}-h(r)}}.
\end{equation}
However, most cases it is not easy to solve this integral. For example in the case of $\frac{\tilde{m}}{r_0} \ll 1$, the deflection angle is found as too small, which is known as the weak lensing. Moreover, $\hat{\alpha}$ grows as $r_0$ closing to the photosphere until it diverges, and become the strong lensing.
\section{Calculation of weak deflection angle of NAT BH}
\subsection{Weak Deflection angle and GBT}
In this section, we calculate the weak deflection angle of NAT BH using the GBT. First, for simplicity, we assume that $\theta=\pi/2$ for equatorial plane and use the spacetime metric given in \ref{metric}, to write the optical metric 
\begin{equation}
    \label{om} 
    dt^2 = \frac{dr^2}{\left(1-\frac{2a_1^2b_1} {r^2}-\frac{2\tilde{m}}{r}\right)^2} + \frac{r^2}{\left(1-\frac{2a_1^2b_1} {r^2}-\frac{2\tilde{m}}{r}\right)} d\phi^2.
\end{equation}
Then we calculate the Gaussian curvature of the optical NAT BH spacetime:
\begin{equation}
    \label{gc}
    K = \frac{R}{2} \approx -2\,{\frac {\tilde{m}}{{r}^{3}}}+6\,{\frac {{ b_1}\, \left( 2\,\tilde{m}-r \right) {
{a_1}}^{2}}{{r}^{5}}}.
\end{equation}
Here, using the above Gaussian curvature of the optical NAT BH spacetime in the GBT, we obtain the deflection angle. The GBT gives relation between the intrinsic geometry of the
spacetime and its topology of the region $D_{R}$ in $M$, with boundary
$\partial D_{R}=\gamma_{\tilde{g}}\cup C_{R}$ \cite{R8}: 
\begin{equation}
\int\limits _{D_{R}}K\,\mathrm{d}S+\oint\limits _{\partial D_{R}}\kappa\,\mathrm{d}t+\sum_{i}\epsilon_{i}=2\pi\chi(D_{R}).
\end{equation}
Note that $\kappa$ is for the geodesic curvature $\kappa=\tilde{g}\,(\nabla_{\dot{\gamma}}\dot{\gamma},\ddot{\gamma})$, where $\tilde{g}(\dot{\gamma},\dot{\gamma})=1$, and the unit
acceleration vector $\ddot{\gamma}$, and $\epsilon_{i}$ corresponds
to the exterior angle at the $i^{th}$ vertex. As $r\rightarrow\infty$,
both jump angles reduces to $\pi/2$, and it is found that $\theta_{O}+\theta_{S}\rightarrow\pi$.
Because of $D_{R}$ is not singular, the Euler characteristic is $\chi(D_{R})=1$. Hence, the GBT is
\begin{equation}
\iint\limits _{D_{R}}K\,\mathrm{d}S+\oint\limits _{\partial D_{R}}\kappa\,\mathrm{d}t+\theta_{i}=2\pi\chi(D_{R}),\label{gaussbonnet}
\end{equation}
in which $\gamma_{\tilde{g}}$ is a geodesic and $\theta_{i}=\pi$ denotes
the total jump angle. Then we have $\kappa(\gamma_{\tilde{g}})=0$.
After recalling the Euler characteristic number, which is $\chi=1$, we find the remaining
part that yields $\kappa(C_{R})=|\nabla_{\dot{C}_{R}}\dot{C}_{R}|$ as
$r\rightarrow\infty$. The radial component of the geodesic curvature
is calculated as follows:
\begin{equation}
\left(\nabla_{\dot{C}_{R}}\dot{C}_{R}\right)^{r}=\dot{C}_{R}^{\varphi}\,\partial_{\varphi}\dot{C}_{R}^{r}+\Gamma_{\varphi\varphi}^{r}\left(\dot{C}_{R}^{\varphi}\right)^{2}.\label{izo1}
\end{equation}
At very large $R$, $C_{R}:=r(\varphi)=r=const$, we have 
\begin{equation}
\left(\nabla_{\dot{C}_{R}^{r}}\dot{C}_{R}^{r}\right)^{r}\rightarrow-\frac{1}{r}.
\end{equation}
It is noted that the geodesic curvature does not depend on topological
defects, $\kappa(C_{R})\rightarrow r^{-1}$. Afterwards, from the optical
spacetime metric  \eqref{om}, one can see that $\mathrm{d}t=r\,\mathrm{d}\,\varphi$, and it follows:
\begin{equation}
\kappa(C_{R})\mathrm{d}t=\mathrm{d}\,\varphi.
\end{equation}
Using the above results, the GBT equation reduces to this form:
\begin{equation}
\iint\limits _{D_{R}}K\,\mathrm{d}S+\oint\limits _{C_{R}}\kappa\,\mathrm{d}t\overset{{r\rightarrow\infty}}{=}\iint\limits _{S_{\infty}}K\,\mathrm{d}S+\int\limits _{0}^{\pi+\hat{\alpha}}\mathrm{d}\varphi. \label{gbtt}
\end{equation}

In the weak deflection limit, one may assume that the light ray is
given by $r(t)=u/\sin\varphi$ at zeroth order. Then we use straight line approximation \cite{R8} as $r = u/ \sin \phi$, where $u$ is the impact parameter, and Eq. \eqref{GBT2} becomes:
\begin{equation}
     \label{allim}
     \hat{\alpha} = - \int_0^\pi \int_{\frac{u}{\sin \phi}}^\infty K dS,
\end{equation}
where $dS=rdr d\phi$.
Note that we ignore the higher order terms. Hence, Eqs. \eqref{gc} and \eqref{allim} are simplified to the following expression for the deflection angle of NAT BH in second order due to the weak lensing:

\begin{equation}
\hat{\alpha}\simeq 3/2\,{\frac {{{ a_1}}^{2}{ b_1}\,\pi}{{u}^{2}}}+4\,{\frac {\tilde{m}}{u}},\label{alpha}
\end{equation}
where the deflection angle is full agreement with the equation (115) in the paper \cite{iz16}. Here, one realizes that depending on the sign of the aether field parameter $b_1$, the light deflection can be more or less than the GR value given by the above first term. For $b_1<0$, the aether field decreases the light deflection angle relative to the Schwarzschild case in GR. This is similar to the effect of charge  in the Reissner-Nordstr{\"o}m solution \cite{Hu, PRD}
for the weak-field limits. Thus, in the presence of $b_1>0$ NAT parameter, the aether increases the deflection angle, and deflection angle reduces to case of Schwarzschild BH when $b_1=0$. The deflection angle in the leading order terms is seen to be in agreement with \cite{iz16}.

\subsection{Weak Deflection angle of NAT BH in a plasma medium}
To take in account of the effects of plasma \cite{Crisnejo:2018uyn}, in this subsection we shall use the case in which light travels from vacuum to a hot, ionized gas medium. Let $v$ be the velocity of light through the plasma. Then, the refractive index $n(r)$ is written as follows:
\begin{equation}
    n(r) \equiv \frac{c}{v} = \frac{1}{dr/dt} \quad \quad\quad\quad\quad \{\because c=1\}.
\end{equation}
Afterwards, we obtain the refractive index $n(r)$ for an NAT BH:
\cite{Crisnejo:2018uyn},
\begin{equation}
    n(r)=\sqrt{1-\frac{\omega_e^2}{{\omega_\infty^2}} \left(1-\frac{2a_1^2b_1} {r^2}-\frac{2\tilde{m}}{r}\right)},
\end{equation}
where $\omega_{e}$ and $\omega_{\infty}$ are the electron plasma frequency and the photon frequency measured by an observer at infinity, respectively. The line-element (\ref{metric}) can be rewritten as:
\begin{widetext}
\begin{equation}
    d \sigma ^ { 2 } = g _ { i j } ^ { \mathrm { opt } } d x ^ { i } d x ^ { j } = \frac { n ^ { 2 } ( r ) } {1-\frac{2a_1^2b_1} {r^2}-\frac{2\tilde{m}}{r}} \left[ \frac{d r ^ { 2 }}{1-\frac{2a_1^2b_1} {r^2}-\frac{2\tilde{m}}{r}} + r^2 d \phi ^ { 2 }\right]. \label{opiz}
\end{equation} 
\end{widetext}
The optical Gaussian curvature becomes:

\begin{eqnarray} 
    K \approx -2\,{\frac {\tilde{m}}{{r}^{3}}}-3\,{\frac {\tilde{m}{\omega_e}^{2}}{{\omega _{\infty}}^{2}{r}^{3}}}+ \left( -6\,{r}^{-4}+12\,{\frac {\tilde{m}}{{r}^{5}}}+
 \left( -10\,{\frac {1}{{\omega _{\infty}}^{2}{r}^{4}}}+52\,{\frac {\tilde{m}
}{{\omega _{\infty}}^{2}{r}^{5}}} \right) {\omega_e}^{2}
 \right) {a_1}^{2}b_1.
\end{eqnarray}

On the other hand, it follows from Eq. (\ref{opiz})
\begin{equation}
\frac{d\sigma}{d\varphi}\bigg|_{C_{r}}=
\sqrt{1-\frac{\omega_e^2}{{\omega_\infty^2}} \left(1-\frac{2a_1^2b_1} {r^2}-\frac{2\tilde{m}}{r}\right)} \left( \frac {  r^2 } { 1-\frac{2a_1^2b_1} {r^2}-\frac{2\tilde{m}}{r}} \right) ^ { 1 / 2 },
\end{equation}
that we have
\begin{equation}
\lim_{R\to\infty} \kappa_g\frac{d\sigma}{d\varphi}\bigg|_{C_R}=1\,.
\end{equation}
For the limit of  $R\to\infty$, and using the straight light approximation $r=u/\sin\varphi$, the GBT becomes \cite{Crisnejo:2018uyn}:
\begin{equation}
\lim_{R\to\infty} \int^{\pi+\alpha}_0 \left[\kappa_g\frac{d\sigma}{d\varphi}\right]\bigg|_{C_R}d\varphi =\pi-\lim_{R\to\infty}\int^\pi_0\int^R_{\frac{u}{\sin\varphi}}\mathcal{K} dS.
\end{equation}

Consequently, the deflection angle yields
\begin{equation}
    \hat{\alpha} \approx 6\,{\frac {\tilde{m}{\omega_e}^{2}}{u{\omega_{\infty}}^{2}}}+4\,{
\frac {\tilde{m}}{u}}+\,{\frac {5{a_1}^{2}b_1{\omega_e}^{2}\pi}{2{u}^{2}{\omega_{\infty}}^{2}}}+\,{\frac {3{a}^{2}b_1\pi}{2{u}^{2}}},\label{pal}
\end{equation}
where the photon rays are moving in a medium of homogeneous plasma. Note that absence of plasma $\left(\omega_{e}=0\right)$,  or $\left(\omega_{e} / \omega_{\infty} \rightarrow 0\right) $ this deflection angle reduces to the vacuum case calculated in \ref{alpha}. It is clear that for the photons propagating in a homogeneous plasma for the case of frequency $\omega_{e} / \omega_{\infty}= 6 \times 10^{-3}$ \cite{BisnovatyiKogan:2010ar}, the deflection angle is increased. But, the effect of the plasma medium can not be detected easily due to its small value in near future observations.

\section{Conclusions}
In this study, we have studied weak gravitational lensing of NAT BH, which is a solution of the new vector-tensor theory. After integrating the deflection angle integral \eqref{allim}, analytically, we have shown that if $\frac{\tilde{m}}{r_0} \ll 1$, the deflection angle gets too small. The latter remark is the evidence of the weak lensing. Remarkably, $\hat{\alpha}$ increases as $r_0$ approaches the photosphere until it diverges to produce the strong lensing. The aether field parameter $b_1$ modifies the gravitational lensing in such a way that when $b_1<0$, the aether field decreases the light deflection angle relative to the Schwarzschild BH of GR. This result is analogue to the effect of charge in the Reissner-Nordstr{\"o}m BH \cite{Hu, PRD} in the weak-field limit. On the other hand, positive NAT parameter increases the deflection angle that reduces to case of Schwarzschild BH when $b_1=0$. Besides, in the existence of plasma $\left(\omega_{e}=0\right)$, the photons propagate in a homogeneous plasma for the case of frequency $\omega_{e} / \omega_{\infty}= 6 \times 10^{-3}$ \cite{BisnovatyiKogan:2010ar, keeton1,keeton2,keeton, XEr,BK11,Turimov} and the deflection angle gets increased. But, the effect of the plasma medium seems not be detected in the near future due its infinitesimal value \cite{izos}. 

As a future work, we plan to proceed to study our topological frame work on the gravitational lensing of rotating NAT BHs, which can be obtained through Newman-Janis algorithm \cite{izfin}, as same as Kerr and/or BTZ which are accepted as more realistic BH geometries. Spraying the particles from their ergosphere effects the moving photons around the rotating BH's photon sphere. Therefore, it will be interesting to analyze the deflection angle of the rotating NAT BH. We believe that the results to be obtained will shed light on future observations.


\end{document}